\documentclass[twocolumn,showpacs,preprintnumbers,amsmath,amssymb]{revtex4}

\usepackage{graphicx} 
\usepackage{dcolumn} 
\usepackage{bm} 

\begin{document}

\title{Spontaneous Spin Polarized Currents in Superconductor-Ferromagnetic 
       Metal Heterostructures}

\author{M. Krawiec}
 \email{m.a.krawiec@bristol.ac.uk}
\author{B. L. Gy\"orffy}
\author{J. F. Annett}
\affiliation{H. H. Wills Physics Laboratory, 
 University of Bristol, Tyndal Ave., Bristol BS8 1TL, UK}

\date{\today}

\begin{abstract}
We study a simple microscopic model for thin, ferromagnetic, metallic layers
on semi-infinite bulk superconductor. We find that for certain values of
the exchange spliting, on the ferromagnetic side, the ground states of such
structures feature spontaneously induced spin polarized currents. Using a
mean-field theory, which is selfconsistent with respect to the pairing
amplitude $\chi$, spin polarization $\vec{m}$ and the spontaneous current 
$\vec{j}_s$, we show that not only there are Andreev bound states in the 
ferromagnet but when their energies $E_n$ are near zero they support 
spontaneous currents parallel to the ferromagnetic-superconducting interface. 
Moreover, we demonstrate that the spin-polarization of these currents 
depends sensitively on the band filling.
\end{abstract}

\maketitle


Recently it has become possible to fabricate high quality interfaces between 
superconductors ($SC$) and a metallic Ferromagnets ($FM$)\cite{Tedrow}. 
Evidently the proximity effect in such $SC$/$FM$ hybrid structures is of 
scientific interest since it facilitates the study of coexistance between 
magnetism and superconductivity \cite{Berk}. Furthermore, it may also become 
technologically important in connection with magnetoelectronics \cite{Bauer} 
and quantum computing \cite{Blatter}. In this letter we report on our study of 
this intriguing phenomenon in a ferromagnetic layers on a bulk superconductor 
with a particular focus on one of its novel physical features, a spontaneously 
induced current in the ground state, which is relavent form both points of view.

Clearly, in the present context the `proximity effect' means both the leakage 
of superconductivity into the non-superconducting, ferromagnetic metal and 
the spin polarization of the superconductor near the interface. In the 
analogous case where the normal metal is non-magnetic ($NM$) this effect has 
been studied for a long time and is, by now, well understood 
\cite{LambertRaimondi}. By contrast, the experimental and theoretical interset 
in the $SC$/$FM$ heterostructures and interfaces is more recent and the subject 
is correspondingly less well developed. Nevertheless, a number of the new 
phenomena, associated with Cooper pairs in an exchange field, have been 
indetified. For instance, it has been found that, as opposed to the $SC$/$NM$ 
case, the pairing amplitude $\chi$ does not decay exponentially to zero on the 
ferromagnetic side of a $SC$/$FM$ interface but oscilates, with a slowly 
decreasing amplitude, as a function of the distance $d$ from the interface 
\cite{Buzdin82}-\cite{Demler}. These oscillations turn out to be manifestations 
of the effect first studied by Fulde and Ferrel \cite{FuldeFerrell} and Larkin 
and Ovchinikov \cite{LarkinOvchinnikov}, often refered to as $FFLO$, and some 
of their more striking consequences, such as the oscillation of the 
superconducting transition temperature T$_{c}$ as a function of the thickness 
of the $FM$ layer has been observed experimentally \cite{Wong}.

Our work is particularly relavent to the predicted \cite{Prokic,Vecino} and 
observed \cite{Kontos} Andereev bound states in clean thin ferromagnetic 
films sandwiched between two superconductors. Interestingly, these states were 
found to form part of the ground or equilibrium states and to give rise to so 
called $\pi$-states of the $SC$/$FM$/$SC$ junction. One of the aim of this 
letter is to argue that a hitherto overlooked salient feature of such $FFLO$ 
$\pi$-jucntion is a spontaneous current parallel to the $FM$/$SC$ interface. We 
shall also show that the spin polarization of such current depends sensitively 
on whether the state has, or has not, particle-hole symmetry.

\begin{figure}[h]
 \resizebox{8.5cm}{!}{
  \includegraphics{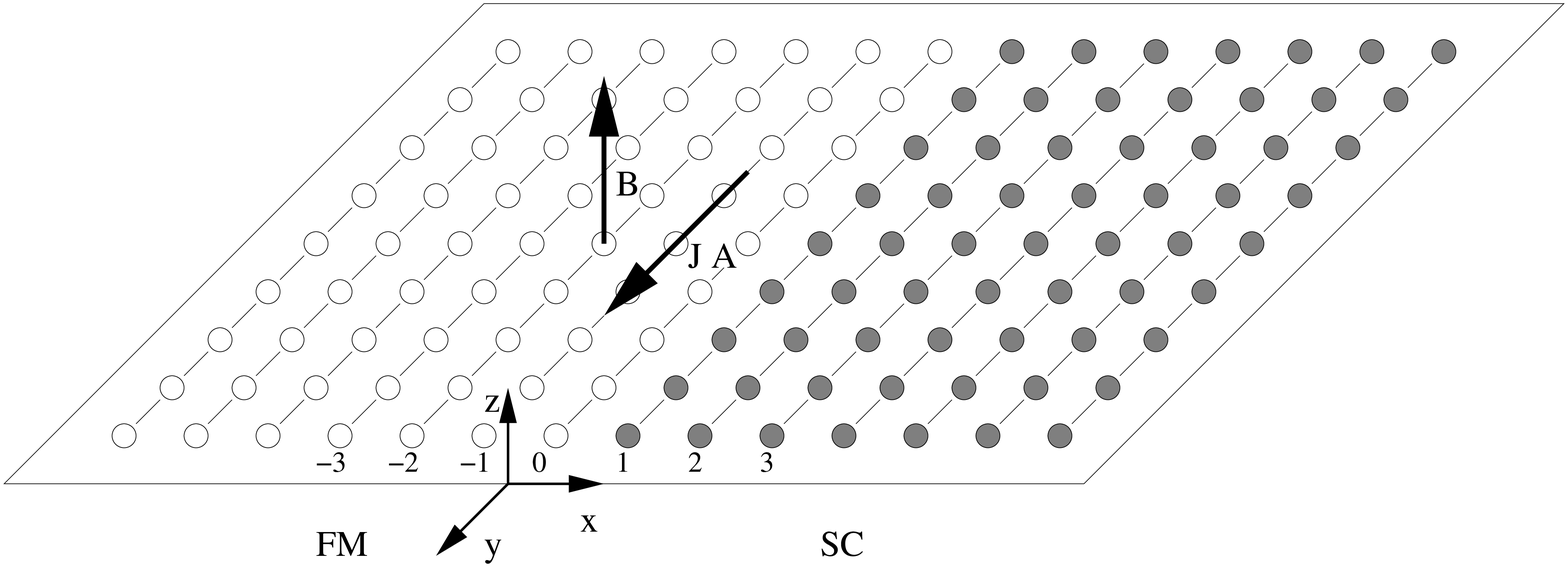}}
 \caption{\label{Fig1} Schematic view of the superconductor-ferromagnet
 interface. Directions of the magnetic field ($B$) as well as vector  potential 
 ($A$) and current ($J$) are indicated.}
\end{figure}

To investigate the occurence spontaneous currents and their polarization in a 
$SC$/$FM$ heterostructure we need a model simple enough to be solved, at least 
in a mean-field approximation, for the magnetization $\vec{m}$, pairing 
amplitude $\chi$ and spontenous current $\vec{j}_s$ self-consistently. Hence, 
for the purpose at hand, we have adopted a single orbital, nearest neighbour 
hopping, negative $U$ Hubbard model for describing the semi-infinite 
superconductor and continued the same Hamiltonian into the ferromagnetic layer 
with the $U$ set equal to zero and the site energies $\varepsilon _{i\sigma}$ 
exchange split. Moreover, we consider the simplest geometry, depicted in 
Fig.\ref{Fig1}, where a magnetic field in one direction, the vector potential 
and a current in another and a spatial modulation in a third orthogonal 
direction can be realised.
Hopefully, while simplifying the calculation this effectively $2D$ system 
will have much in common with its $3D$ counterpart such as a layer of 
ferromagnetic metal deposited on a superconducting substrate 
\cite{Tedrow, Wong} and corresponding sandwich structures \cite{Ryazanov}.

In short, our model Hamiltonian is given by
\begin{eqnarray}
 H =  \sum_{ij\sigma} [t_{ij} + (\varepsilon_{i\sigma} - \mu) \delta_{ij}] 
      c^+_{i\sigma} c_{j\sigma} + 
      \sum_{i\sigma} \frac{U_i}{2} \hat n_{i\sigma} \hat n_{i-\sigma}
 \label{Hamiltonian}
\end{eqnarray}
where, in the presence of a vector potential $\vec{A}(\vec{r})$, 
$t_{ij} = - t e^{-i e \int_{\vec{r}_i}^{\vec{r}_{j}} 
\vec{A}(\vec{r}) \cdot d\vec{r}}$ for nearst neighbour lattice sites at 
spatial positions $\vec{R}_i$ and $\vec{R}_j$, the site energies 
$\varepsilon_{i\sigma}$ are $0$ on the superconducting side and equal to 
$\frac{1}{2} E _{ex}\sigma$ on the ferromagnetic side, $\mu$ is the chemical 
potential, $U_i$ is $U_S < 0$ in the superconductor and zero elswhere,
$c^+_{i\sigma}$, ($c_{i\sigma}$) are the usual electron creation (annihilation) 
operators and $\hat n_{i\sigma} = c^+_{i\sigma} c_{i\sigma}$. Note that the 
above description of the electrons with charges $e$ includes a coupling to a 
magnetic field $\vec{B}(\vec{r}) = \vec{\nabla} \times \vec{A}(\vec{r})$. 
Evidently such provision will be necessary for calculating the effects of 
currents on the electronic states.

In what follows we shall study the above model in the 
Spin-Polarized-Hartree-Fock-Gorkov ($SPHFG$) approximation. We shall work in 
the Landau gauge where $\vec{B}=(0,0,B_z(x))$ and hence $\vec{A}=(0,A_y(x),0)$. 
Furthermore, we assume that the effective $SPHFG$ Hamiltonian is periodic in 
the direction parallel to the inteface and therefore we work in $\vec{k}$ 
space in the $y$ direction but in real space in the $x$-direction (see
Fig.\ref{Fig1}). Labeling the planes in Fig.\ref{Fig1} by integers $n$ and $m$ 
at each $k_y$ point of the Brillouin zone we shall solve the following $SPHFG$ 
Nambu, spin ($\alpha$) and layer index ($n$) matrix equation:
\begin{eqnarray}
 \sum_{m',\gamma,k_y} H^{\alpha\gamma}_{nm'}(\omega,k_y) 
 G^{\gamma\beta}_{m'm}(\omega,k_y) =
 \delta_{nm} \delta_{\alpha\beta}
 \label{HFG}
\end{eqnarray}
where the only non-zero elements are:
$H^{11}_{nm}$ and $H^{22}_{nm} = (\omega - \frac{1}{2} \sigma E_{ex} \pm \mu 
 \pm  t cos(k_y \mp eA(n)))\delta_{nm} \pm t \delta_{n,n+1}$ for the upper and 
lower sign respectively, 
$H^{33}_{nm} = H^{11}_{nm}$ and $H^{44}_{nm} = H^{22}_{nm}$ with $\sigma$ 
replaced by $-\sigma$ in both cases, 
$H^{12}_{nm} = H^{21}_{nm} = - H^{34}_{nm} = - H^{43}_{nm} = 
\Delta_n \delta_{nm}$ 
and $G^{\alpha\beta}_{nm}$ is corresponding retarded Green's function ($GF$).

As usual, selfconsistency is assured by the relation:
\begin{eqnarray}
 \Delta_n = U_n \sum_{k_y} 
 \langle c_{n\downarrow}(k_y) c_{n\uparrow}(k_y) \rangle = 
 \nonumber\\
 - U_n \sum_{ky} \int d\omega \frac{1}{\pi} 
 {\rm Im} G^{12}_{nn}(\omega,k_y) f(\omega)
 \label{Delta}
\end{eqnarray}
where $f(\omega)$ is the Fermi distribution function. Moreover, the $FM$ order 
parameter is given by
\begin{eqnarray}
 m_n = \frac{1}{2} (n_{n\uparrow} - n_{n\downarrow}) = 
 - \frac{1}{2\pi} \sum_{ky} \times
 \nonumber\\
 \int d\omega 
 {\rm Im} (G^{11}_{nn}(\omega,k_y) - G^{33}_{nn}(\omega,k_y)) 
 f(\omega)
 \label{m}
\end{eqnarray}

Since our model includes a  vector potential the solution of Eq.\ref{HFG} will 
imply a current. For spin up electrons, in the $y$-direction this can be 
calculated from the relation:
\begin{eqnarray}
 J_{y\uparrow (\downarrow)}(n) = - 2 e t \sum_{k_y} sin(k_y - e A_y(n)) 
 \times
 \nonumber\\
 \int d\omega \frac{1}{\pi} {\rm Im} G^{11(33)}_{nn}(\omega,k_y) f(\omega)
 \label{current}
\end{eqnarray}
which follows from the continuity equation for the charge.

Finally, the above current will give rise to a vector potential 
$A_{new}(\vec{r})$  which will have to be used to update $A(\vec{r})$ in 
Eq.\ref{HFG} at the end of each selfconsistency cycle. We calulated this new 
vector potential by solving numerically Ampere's law, 
$\frac{d^2 A_y(x)}{d x^2} = - 4 \pi J_y(x)$, which for the lattice problem at 
hand, is 
\begin{eqnarray}
 A_y(n+1) - 2 A_y(n) + A_y(n-1) = - 4 \pi J_y(n)
 \label{Maxwell}
\end{eqnarray}

We have solved Eqs.\ref{Delta}-\ref{Maxwell} using an appropriately simplified 
`principal layers method' \cite{Turek} which we shall describe elswhere 
\cite{KGA}. The rest of this letter is a brief summary of our resuls.

Firstly, since we determined the order parameters, $\chi_n$ and $m_n$, 
on both sides of the inerface fully self-consistently, we were able to study 
both a superconducting and a magnetic proximity effects. Although $\Delta_n=0$ 
on the ferromagnetic side, due to the fact that $U=0$, the pairing amplitude 
$\chi_n= \langle c_{n\downarrow} c_{n\uparrow} \rangle$ does not have to be, 
and indeed it turns out not to be, zero. Given the well understood effect of 
the exchange field in a bulk superconductor 
\cite{LarkinOvchinnikov,FuldeFerrell} it is not all together surprising that 
we find that on entering into the ferromagnet $\chi_n$ oscillates as a 
function of the distance from the interface. In fact our numerical results fit 
the analytic formula $\chi(x) \sim sin(x/\xi_F)/(x/\xi_F)$, where 
$\xi_F= t / E_{ex}$ is the ferromagnetic coherence length, and hence are fully 
consisitent with those of Ref.\cite{Demler,Vecino,Halterman,Bagrets}.
As it is well known at a $NM$/$SC$ inerface $\chi_n$ decays exponentially in 
the normal region with the decay length $\xi_S$. Thus, although the 
supercoducting coherence length $\xi_S >> \xi_F$ due to the functional form, 
polynomial as opposed to exponential, of the decay $\chi_n$ enters more deeply 
into the ferromagnet then into the normal metal. An other novelty of the 
$FM$/$SC$ interface, when compared with its well studied  $NM$/$SC$ 
counterpart, is the entry of magnetism into the superconductor. We find that 
the spin polarization does not oscillate but decays monotonically on a length 
scale of order of the $\xi_F$.
\begin{figure}[h]
 \resizebox{8cm}{4.8cm}{
  \includegraphics{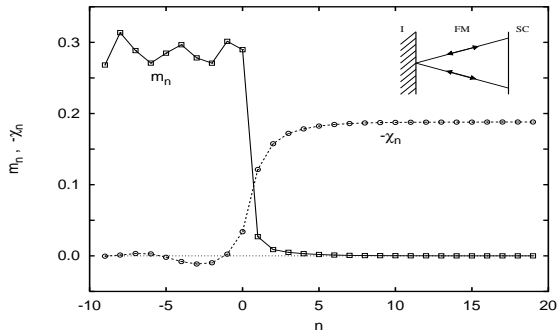}}
 \caption{\label{Fig2} Magnetization and the pairing amplitude as a function of
 the distance from the interface for $E_{ex} = 1.273$ and $U_S = -2$, which
 gives $\Delta_S = 0.376$ in units of the hopping integral $t$. Note that we
 plot $-\chi_n$ which corresponds to positive $\Delta_n$ for $U_n > 0$. Inset:
 Example of the semiclassical paths corresponding to Andreev bound states.}
\end{figure}
As shown in Fig.\ref{Fig2}, somewhat suprisingly, $m_n$ oscillates on the
ferromagnetic side but dacays exponentially on the superconducting side of our 
hybrid structure. This behavior is in full agreement with the results of 
Demler et al. \cite{Demler}.

\begin{figure}[h]
 \resizebox{8cm}{4.8cm}{
  \includegraphics{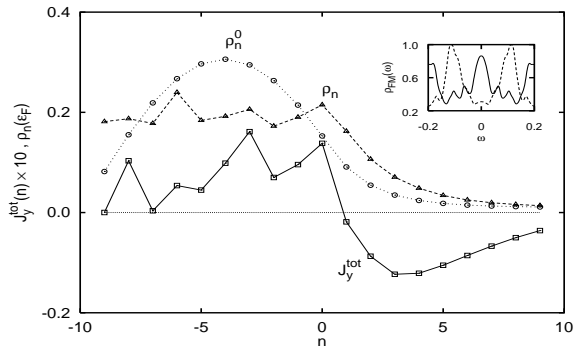}}
 \caption{\label{Fig3} Total current 
 $J^{tot}_y(n) = J_{y\uparrow}(n) + J_{y\downarrow}(n)$ and the total density 
 of states at the Fermi level $\rho_n(\varepsilon_F)$. For comparison the $DOS$ 
 for $E_{ex}=0$ ($\rho^0_n(\varepsilon_F)$) is also shown. Inset: Integrated 
 over the $FM$ side total $DOS$ $\rho_{FM}(\omega)$. Dashed (solid) line 
 corresponds to the solution with (without) spontaneous current. The parameters 
 are the same as in the Fig.\ref{Fig2}}
\end{figure}

Of course, the most remarkable feature of the above solution is that the 
iterations of the $SPHFG$ equations frequently converge to a finite value of 
the current $j_y(n)$ even though the external vector potential is zero. We have 
checked that the exsistence of such spontaneous current lowers the energy of 
the system. As shown in Fig.\ref{Fig3}, it flows in the positive $y$ direction 
on the ferromagnetic side, and in negative in the superconductor and sums, 
reassuringly, to zero over all layers. We have also found numerically, that 
there is a magnetic flux $\Phi \approx \Phi_0/2$, $\Phi_0$, being the flux 
quantum, associated with this current distribution.

To shed light on the origin of the above spontaneous current, in 
Fig.\ref{Fig3}, we also show the layer resolved density of states, namely the 
spectral function $\rho_n(\varepsilon_F) = -\frac{1}{\pi}\sum_{k_y} 
{\rm Im} (G^{11}_{nn}(\varepsilon_F,k_y) + G^{33}_{nn}(\varepsilon_F,k_y))$, 
at the Fermi energy $\varepsilon_F$. Clearly, the oscillations of the layer 
resolved current tracks those of the density of states. Further insight follows 
if we compare this with the corresponing quantity, denoted by 
$\rho^0_n(\varepsilon_F)$ and represented by the dot line, in the case where 
the exchange spliting, $E_{ex}$, is set equal to zero. Since the rise and fall 
of $\rho^0_n(\varepsilon_F)$ across the ferromagnetic layer can be readily 
inetrpreted as the order parameter amplitude of an Andreev bound state,
corresponding to the semiclassical path depicted in the inset of the
Fig.\ref{Fig2}, we can regard $\chi_n$ in Fig.\ref{Fig2} and 
$\rho_n(\varepsilon_F)$ in Fig.\ref{Fig3} as an indication that a similar bound 
state is formed in the much more complicated case of finite exchange field. In 
fact we can indentify such $FFLO$-Andreev bound states as peaks in the full 
quasiparticle density of states for the ferromagnetic layers. As might be 
expected these are exchange split and move around as a function of the exchange 
field $E_{ex}$. Investigating the correlation between such bound states and the 
current carrying capacity of a solution we find that in the ground state a 
current flows only when the energy of one of the $FFLO$-Andreev bound state 
is near the Fermi energy. Furthermore, in the presence of the current, this 
state splits, thus lowering the total energy of the system. An example of such 
zero energy bound state is depicted in the inset of the Fig.\ref{Fig3}.

\begin{figure}[h]
 \resizebox{8cm}{4.8cm}{
  \includegraphics{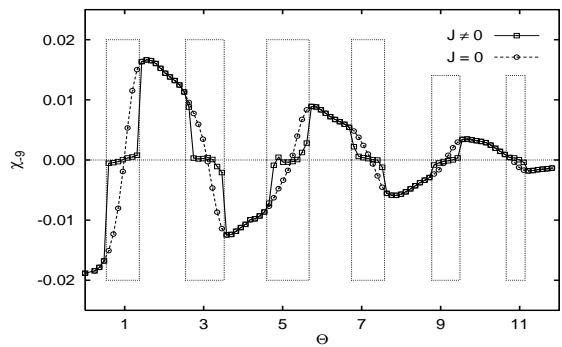}}
 \caption{\label{Fig4} Pairing amplitude at the surface of the $FM$ layer
 oposite to superconductor ($n = - 9$) as a function of the dimensionless 
 parameter $\Theta = 3 d E_{ex} / \pi t$. Note pinning of the $\chi$ to zero as 
 long as current flows.}
\end{figure}

Zero energy Andeev bound states in $SC$/$NM$/$SC$ juctions usually lead to a so 
called $\pi$-states of the two superconductor in which the phases of their 
order parameters differ by $\pi$. Recently, Chtchelkatchev et al.
\cite{Chtchelkatchev} suggested that the same is true for a $SC$/$FM$/$SC$ 
junction in the presence of fully developed $FFLO$ phenomena. Interestingly, 
although the structure we have been studying has only one superconducting 
region it turns out to display properties analougous to those of such 
$\pi$-junctions. To see these we studied the order parameter $\chi_{-9}$ at the 
surface of the ferromagnetic layer opposite to the superconductor. Since 
$\chi_n$ is negative in superconductor when $\chi_{-9}$ is positive we may
describe the system as being in a $\pi$-state. In Fig.\ref{Fig4} we display 
our results for $\chi_{-9}$ as a function of the dimensionless exchange field 
$\Theta = 3 d E_{ex} / \pi t$ ($d$ is the number of $FM$ layers).
Evidently, there are regions of $\Theta$ for which the sytem can be said to be 
in $\pi$-state, in close agreement with the analougous results of 
Chtchelkatchev et al. \cite{Chtchelkatchev}.

\begin{figure}[h]
 \resizebox{8cm}{4.8cm}{
  \includegraphics{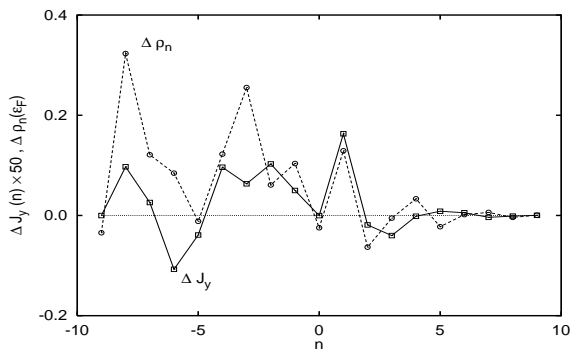}}
 \caption{\label{Fig5} Spin polarized current 
 $\Delta J_y (n) = J_{y\uparrow}(n) - J_{y\downarrow}(n)$ and the difference of
 the spin up and spin down density of states at the Fermi level 
 $\Delta \rho_n(\varepsilon_F) = 
 \rho_{n\uparrow }(\varepsilon_F)-\rho_{n\downarrow}(\varepsilon_F)$
 as a function of the distance from the interface. Here the band filling is
 $0.636$.}
\end{figure}

Remarkably, the states with spontaneous currents correspond to regions of the
exchange field $\Theta$ where the order parameter $\chi_{-9}$ is near zero. As 
can be easily read off from these plots, in these regions the presence or 
absence of the currents have a dramatic effect on the order parameter. 
Evidently, the spontaneous current pins the order parameter on the 
non-superconducting side of the ferromagnetic layer ($\chi_{-9}$) to zero. This 
appears to stablise the zero energy $FFLO$-Andreev bound state.
Given their physical origin one might expect the above sponteneous currents to 
be spin polarised. This is indeed the case. In fact, we find that the degree 
of spin polarization is largely detemined by the difference in the spin up and 
spin down densities of state at the Fermi energy: 
$\Delta \rho_n = 
 \rho_{n\uparrow }(\varepsilon_F) - \rho_{n\downarrow}(\varepsilon_F)$. 
In the above calculations we have assumed a half filled band, that is to say 
particle hole symetry, and hence we have found no difference between the spin 
up and spin down currents. However, further calculations, away from particle 
hole symetry, revealed much larger spin polarizations. An example of this is 
reported in Fig.\ref{Fig5}.

In summary, we have demonstrated that zero energy Andreev bound states in
ferromagnetic layers deposited on a superconducting substrate will carry
currents in the ground state of such a hybrid structure. Moreover, we found 
that such states will form only in certain regions of the exchange field 
$E_{ex}$ and layer thicknes $d$ phase diagram. In particular, we investigated 
the cases $0 < E_{ex}/t < 3$,  $0 < d < 20$. Finally we found that the spin 
polarization of the current is closely related to the difference in the spin up 
and spin down density of states at the Fermi level.

This work has been supported by Computational Magnetoelectronics Research
Training Network under Contract No. HPRN-CT-2000-00143.


\end{document}